\documentclass[twocolumn,showpacs,superscriptaddress,amsmath,amssymb]{revtex4-1}
\usepackage{graphicx}
\usepackage{dcolumn}
\usepackage{bm}
\usepackage{color}
\usepackage{epsfig}
\usepackage{epstopdf}
\usepackage{mathrsfs}
\usepackage{hyperref}
\usepackage{amsmath}
\usepackage{amsxtra}
\usepackage{amsfonts}

\usepackage{multirow}
\usepackage{makecell}
\usepackage{longtable}
\usepackage{diagbox}
\usepackage{slashbox}
\usepackage{url}

\def\bd{
\begin{document}} \def\ed{\end{document}}
\def\bmp{\begin{minipage}} \def\emp{\end{minipage}}
\def\bcc{\begin{center}} \def\ecc{\end{center}}     \def\npg{\newpage}
\def\beq{\begin{equation}} \def\eeq{\end{equation}} \def\hph{\hphantom}
\def\be{\begin{equation}} \def\ee{\end{equation}} \def\r#1{$^{[#1]}$}
\def\n{\noindent} \def\ni{\noindent} \def\pa{\parindent}
\def\hs{\hskip} \def\vs{\vskip} \def\hf{\hfill} \def\ej{\vfill\eject}
\def\cl{\centerline} \def\ob{\obeylines}  \def\ls{\leftskip}
\def\underbar#1{$\setbox0=\hbox{#1} \dp0=1.5pt \mathsurround=0pt
   \underline{\box0}$}   \def\ub{\underbar}    \def\ul{\underline}
\def\f{\left} \def\g{\right} \def\e{{\rm e}} \def\o{\over} \def\d{{\rm d}}
\def\vf{\varphi} \def\pl{\partial} \def\cov{{\rm cov}} \def\ch{{\rm ch}}
\def\la{\langle} \def\ra{\rangle} \def\EE{e$^+$e$^-$} \def\pt{p_{\rm t}}
\def\pti{p_{{\rm t},i}} \def\vti{v_{{\rm t},i}}
\def\ptj{p_{{\rm t},j}}\def\Pt{P_{\rm t}} \def\vt{v_{\rm t}}

\def\bitz{\begin{itemize}} \def\eitz{\end{itemize}}
\def\btbl{\begin{tabular}} \def\etbl{\end{tabular}}
\def\btbb{\begin{tabbing}} \def\etbb{\end{tabbing}}
\def\beqar{\begin{eqnarray}} \def\eeqar{\end{eqnarray}}
\def\\{\hfill\break} \def\dit{\item{-}} \def\i{\item}
\def\bbb{\begin{thebibliography}{9} \itemsep=-1mm}
\def\ebb{\end{thebibliography}} \def\bb{\bibitem}
\def\bpic{\begin{picture}(260,240)} \def\epic{\end{picture}}
\def\akgt{\cl{\bf ACKNOWLEDGMENTS}}
\def\fgn{\noindent{\bf\large\bf figure captions}}
\def\m1{\langle N_p\rangle} \def\u2{\langle N_{\bar p}\rangle} \def\Nap{N_{\bar
p}}
\def\lan{\langle}
\def\ran{\rangle}
\def\p{\pi}
\def\ifmath#1{\relax\ifmmode #1\else $#1$\fi}%
\def\rc{\ifmath{{\mathrm{c}}}}
\def\cut{\ifmath{{\mathrm{cut}}}}
\def\rF{\ifmath{{\mathrm{F}}}}
\def\rK{\ifmath{{\mathrm{K}}}}
\def\rp{\ifmath{{\mathrm{p}}}}
\def\rt{\ifmath{{\mathrm{t}}}}
\def\LAB{\ifmath{{\mathrm{LAB}}}}
\def\cut{\ifmath{{\mathrm{cut}}}}
\def\beq{\begin{equation}}
\def\eeq{\end{equation}}

\newcommand{\cinst}[2]{$^{\mathrm{#1}}$~#2\par}
\newcommand{\crefi}[1]{$^{\mathrm{#1}}$}
\newcommand{\crefii}[2]{$^{\mathrm{#1,#2}}$}
\newcommand{\crefiii}[3]{$^{\mathrm{#1,#2,#3}}$}
\newcommand{\HRule}{\rule{0.5\linewidth}{0.5mm}}
\newcommand{\minitab}[2][l]{\begin{tabular}{#1}#2\end{tabular}}

\bd
\title{Locating fixed points in the phase plane}
\author{Yanhua Zhang}
\affiliation{Key Laboratory of Quark and Lepton Physics (MOE) and
Institute of Particle Physics, Central China Normal University, Wuhan 430079, China}
\author{Yeyin Zhao}
\affiliation{Key Laboratory of Quark and Lepton Physics (MOE) and
Institute of Particle Physics, Central China Normal University, Wuhan 430079, China}
\author{Lizhu Chen}
\affiliation{School of Physics and Optoelectronic Engineering, Nanjing University of Information Science and Technology, Nanjing 210044, China}
\author{Xue Pan}
\affiliation{School of Electronic Engineering, Chengdu Technological University, Chengdu 611730, China}
\author{Mingmei Xu}
\affiliation{Key Laboratory of Quark and Lepton Physics (MOE) and
Institute of Particle Physics, Central China Normal University, Wuhan 430079, China}
\author{Zhiming Li}
\affiliation{Key Laboratory of Quark and Lepton Physics (MOE) and
Institute of Particle Physics, Central China Normal University, Wuhan 430079, China}
\author{Yu Zhou}
\affiliation{ No.1 Middle School Affiliated to Central China Normal University,  Wuhan 430223, China}
\author{Yuanfang Wu}\email{wuyf@mail.ccnu.edu.cn}
\affiliation{Key Laboratory of Quark and Lepton Physics (MOE) and
Institute of Particle Physics, Central China Normal University, Wuhan 430079, China}

\begin{abstract}
The critical point is a fixed point in finite-size scaling. To quantify the behaviour of such a fixed point, we define, at a given temperature and scaling exponent ratio, the width of scaled observables for different sizes. The minimum of the width reveals the position of fixed point, its corresponding phase transition temperature, and scaling exponent ratio. The value of this ratio tells the nature of fixed point, which can be a critical point, a point of the first order phase transition line, or a point of the crossover region. To demonstrate the effectiveness of this method, we apply it to three typical samples produced by the 3D three-state Potts model. Results show the method to be more precise and effective than conventional methods. Finally, we discuss a possible application at the beam energy scan program of relativistic heavy-Ion collision.
\end{abstract}


\maketitle
\section{Introduction}

The phase diagram in statistical physics labels the boundary where phase transition (PT) happens.  In order to draw this boundary with precision, some key points of the boundary line should be located firstly. The critical point (CP) of the second-order PT is one of these points. Its temperature, so called critical temperature ($T_{\rm C}$), is one of the most interesting parameters~\cite{Nucl-1, Nucl-2, Percolation-1, Percolation-2, STAR-BES, STAR-c, Fisch-QCD, Luo-Xu-Nature}.

As we know, if the volume of the system is large enough, $T_{\rm C}$ can be approximately determined by the peak position of the distribution of susceptibility~\cite{sus-2,sus-3,Newman}. While, for the system with finite volume, $T_{\rm C}$ has to be determined by the finite-size scaling (FSS)~\cite{Nucl-1, Nucl-2,book-1,book-2, RoyPRL, obversables-scaling-b}.

The FSS of observable $Q(T,L)$ has usually the form~\cite{book-1,book-2},
\begin{equation}\label{scaling function 1}
Q(T,L)=L^{-\lambda/\nu}f_{Q}(t L^{1/\nu}).
\end{equation}
\noindent Where $t=(T-T_C)/T_C $ is reduced temperature. $f_{Q}$ is scaling function. $t L^{1/\nu}$ is scaled variable. $\lambda$ and $\nu$ are respectively the scaling exponents of observable and the correlation length $\xi\propto |\tau|^{-\nu}$. These exponents characterize the universal class of PT. Here the scaling exponent ratio (SER) $a=\lambda/\nu$ is {\it a fraction} between the spatial dimension $d$ and zero.

The FSS means that the observables for different system sizes and temperatures can be scaled to an universal scaling function, as showed in Fig.~1(a). There are three parameters, $T_{\rm C}$, $\nu$, and $\lambda$ (or $a$), in this scaling function. To determine $T_{\rm C}$ and scaling exponents, it is usually assumed that the observables fall in a certain universal class. $T_{\rm C}$ is obtained by the Binder cumulant ratio, and scaling exponents are obtained by the best scaling plot of the observables of different sizes~\cite{Binder-a,Binder-b,potts-ising-b}. Selection of the best scaling is done through visual observation. The method, therefore, often lack in precision~\cite{Nature333-Lutz-Warnke}.

On the other hand, the FSS also implies a {\it fixed point} (FP). If we plot scaled observable versus $T$, instead of scaled variable $t L^{1/\nu}$, the curves of different sizes will intersect at the FP, where $T=T_{\rm C}$ and scaling function $f_{Q}(0)$ is independent of system size, i.e., Fig.~1(b). The parameter $\nu$ is therefore omitted in this plot.

\begin{figure}
\centering
\includegraphics[width=1\linewidth]{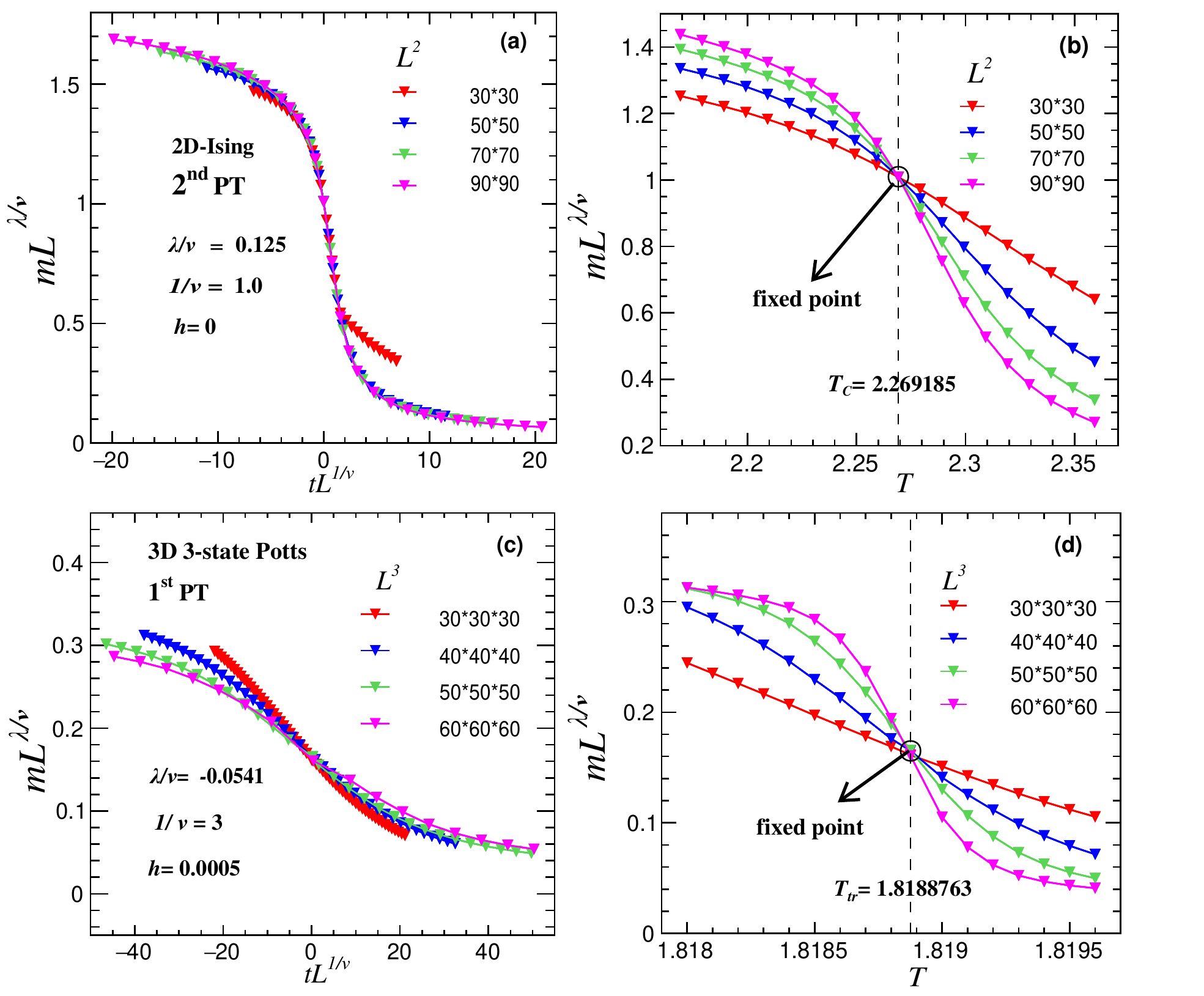}
\caption{\label{Fig. 1} The scaled order parameter ($m L^{\lambda/\nu}$) versus scaled variable ($tL^{1/\nu}$, left column (a) and (c)), and $T$ (right column (b) and (d)) for the second (top panels, 2D Ising model with external field $h=0$) and the first-order PTs (bottom panels, 3D three-state Potts model with $h=0.0005$), respectively.}
\end{figure}

The FP is obtained from the scale transformation of renormalization group~\cite{book-2, Fisher-1974,Wilson-1974}. In phase diagram, not only the CP is a FP, but also the point of the first order PT line~\cite{Fisher:1982xt,vanLeeuwen:1975zz, PhysRevLett.35.477,book-2}, cf., Fig.~1(c) and (d). The SER $a$ of  the first-order PT line is {\it an integer}, in contrast to that of the second-order PT, where $a$ is a fraction.

It is clear that the scaling holds in a limited range of system size or temperature, as showed in Fig.~1(a) and (c). The valid range of size or temperature varies with studying system, its spatial dimension, and the order of PT. For a relatively small size or a substantial deviation from the critical temperature, the scaling is violated. So with the change of size or temperature, there should be a correction for the scaling.
However, as the scaled variable is the product of reduced temperature and size, the temperature and size are tied together in the scaling plot. The influence of size or temperature is therefore difficult to quantify.

In contrast to the scaling plot in Fig.~1(a) or (c), the plot of FP in Fig.~1(b) or (d) shows clearly how scaled observables for different sizes vary with temperature. Any deviation from the critical (or transitional) temperature, the curves for different sizes are separate from each other. They converge, or intersect only at the critical (or transitional) temperature. This feature allows us to quantify the deviation for different sizes at a given temperature, and to accurately locate critical temperature.

At a given $T$ and $a$, all points of curves for different sizes can be regarded as a set. The width of the set can be defined as the square root of the variance of scaled observables. Usually, the defined width depends on $T$ and $a$. When $T$ and $a$ both approach transition values, all points in the set overlap within error, i.e., FP,  and the defined width reaches its minimum. So defined width well quantifies the behaviour of FP. $T_{\rm FP}$ and $a_{\rm FP}$ can be determined by the minimization of the width.

For the crossover region, observable is independent of system size in the region of transition temperature~\cite{Fodor-Nature}. This character can be generalized to SER $a=0$ in scaling form Eq.~(\ref{scaling function 1}. It implies that curves for different sizes, as those showed in Fig.~1(b) or (d), overlap within errors at the region of transition temperature. The defined width is therefore also minimized, same as the case of FP.

Therefore, in general, the value of $a_{\rm FP}$ reveals the nature of FP, which is either a CP, a point of the first-order PT line, or a point of crossover region.

In this paper, we first quantify the behaviour of FP in section II. Then, in section III, three samples of order parameter are produced by the 3D three-states Potts model, where the CP, the point of the first-order PT line, and the point of the crossover region are all well defined. In section IV,  we demonstrate that in the plane of $T$ and $a$, the contour plot of defined width precisely locates the position of FP in three given samples, respectively. In section V, a possible application of the method at the RHIC BES is discussed. Finally, a summary is given in section VI.

\section{Description of fixed point}

In Fig.~1(b) and (d), at a given $T$, when $T$ approaches $T_{\rm C}$ (or $T_{tr}$) in value, the points of curves for different sizes approach each other, allowing the formation of an intersection point.
When $T$ deviates from $T_{\rm C}$ (or $T_{tr}$), all points separate from each other. In order to quantify the relative distance between the points, we define, at a given $a$, the width of all size scaled observables ($Q(T,L)L^a$) to be square root of their variance,  i.e.,
\begin{flalign}\label{define D}
 D(T,a)=\sqrt{\frac{\Delta S_{Q(T,L)L^a}}{N_L-1}}.
\end{flalign}
Here, $D(T,a)$ varies with $T$ and $a$. $N_L$ is the number of sizes. $\Delta S_{Q(T,L)L^{a}}$ is the error weighted variance of all scaled observables to their mean position, i.e.,
\begin{flalign}\label{define D 1}
 \hspace{-5mm}     \Delta S_{Q(T,L)L^{a}}=\sum_{i=1}^{N_L}\frac{[Q(T,L_i)L^{a}_i-\langle Q(T,L)L^{a}\rangle]^2}{\omega^2_i}.
\end{flalign}
$\omega_i = \delta [Q(T,L_i)L^{a}_i]$ is the error of $Q(T,L_i)L^{a}_i$. $\langle Q(T,L)L^{a}\rangle$ is the weighted mean, i.e.,
\begin{flalign}\label{define D 2}
\langle Q(T,L)L^{a}\rangle = \frac{\sum_{i=1}^{N_L}Q(T,L_i)L^{a}_i /\omega^2_i}{\sum_{i=1}^{N_L}1/\omega^2_i}.
\end{flalign}

$D(T,a)$ describes the relative distance of all points to their mean position. When $T$ and $a$ deviate from critical (transitional) values, the points of curves for different sizes move away from each other, and the value of $D(T,a)$ increases. When $T$ and $a$ are both the critical (transitional) values, all points overlap within error. $D(T,a)$ reaches its minimum, around unity.

Such defined width is analogous to $\chi^2$ of curve-fitting, where the variance is the measured points to a given curve. The minimum of $\chi^2$ corresponds to the best curve-fitting.

In a real experiment, due to the error of observable and other uncertainties, FP may not be an ideal point, and the minimum of $D(T,a)$ may be larger than unity. Nevertheless, if there is a FP in $T$ and $a$ plane, the $D(T,a)$ will change with $T$ and $a$ and converge to a minimum. This converging trend of  $D(T,a)$ is essential for forming a FP.

\begin{figure}[!htb]
\centering
\includegraphics[width=1\linewidth]{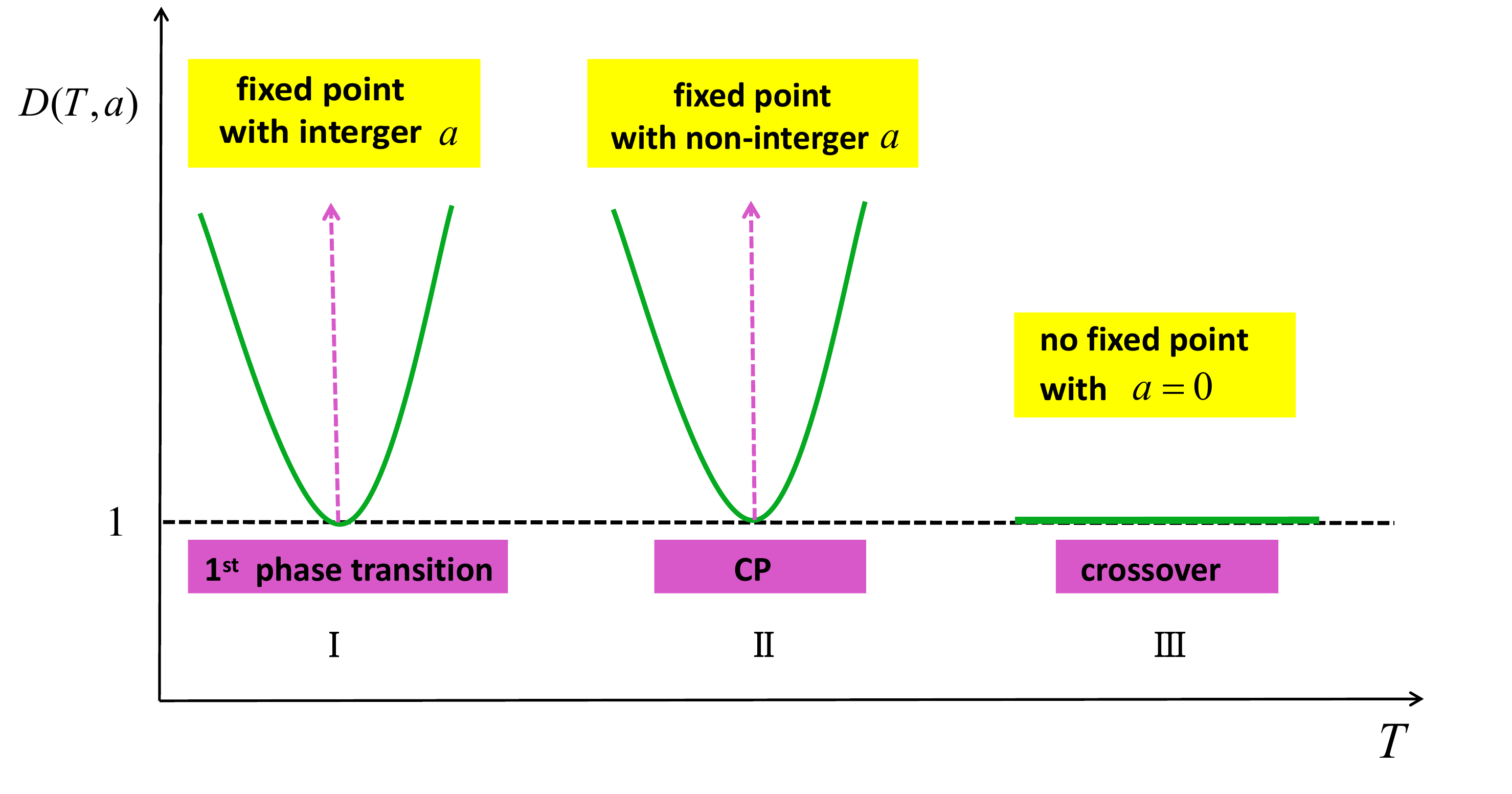}
\caption{\label{Fig. 2} (Color line) $D(T,a)$ nearby the temperatures of the first (I),  second (II)-order PT, and the crossover region (III).}
\end{figure}

At three different regions of phase plane, $D(T,a)$, $T$ and $a$ are expected to be the cases of I, II and III, respectively, as showed in Fig.~2.

In the case of I, $T$ is low. $D(T,a)$ has a minimum at the transitional $T$, where the SER $a$ is an integer. It characterizes the FP of the first-order PT.

In the case of II,  $T$ is in the middle. $D(T,a)$ has also a minimum at the $T_{\rm C}$, and the SER $a$ is a fraction, in contrast to the case of I. It characterizes the FP of the second-order PT, i.e., CP.

In the case of III, $T$ is high. $D(T,a)$ is a constant, independent of $T$, and the SER $a$ is zero. The observable is independent of system size. It characterizes the FP of the crossover region.

In the following, we will show in practice how to locate FP and determine its $T_{\rm FP}$ and $a_{\rm FP}$.

\section{Three samples of the Potts model}

The same as QCD with finite temperature and infinite quark mass, the 3D three-state Potts model has Z(3) global symmetry~\cite{potts-Z-a,potts-Z-b,potts-Z-c,potts-Z-d}.
The external magnetic field in the Potts model plays the role of quark mass in the finite
temperature QCD. At vanishing external field, the temperature-driven PT has
been proved to be of the first-order~\cite{potts-first-order-a, potts-first-order-b}. As external field increases, the first-order PT weakens and ends at the CP
$(1/T_{\rm C}, h_{\rm C}) = (0.54938(2), 0.000775(10))$~\cite{potts-Z-b,potts-ising-b},
which belongs to 3D Ising universality class, the same as deconfining and chiral CPs in QCD.
Above the critical temperature, the transition is a crossover~\cite{Fodor-Nature}.

The 3D three-state Potts model is described in terms of spin variable
$s_i \in {1, 2, 3}$, which is located at sites $i$ of a cubic lattice with size $V = L^3$.
The Hamiltonian of the model is defined by~\cite{Xue-JPG42,potts-ising-b},
\begin{flalign}\label{H}
H=\beta E-hM.
\end{flalign}
\noindent The partition function is,
\begin{flalign}\label{Z}
Z(\beta, h) =\sum_{\{{s}_{i}\}}e^{-(\beta E-hM)}.
\end{flalign}
Where $\beta=1/T$ is reciprocal of temperature, and $h=\beta H$ is normalized external magnetic field. $E$ and $M$ denote the energy and magnetization respectively, i.e.,
\begin{flalign}\label{E and M}
E=-J\sum_{\langle i,j\rangle}\delta(s_i,s_j), \ {\rm and} \ M=\sum_{i}\delta(s_i,s_g).
\end{flalign}
Here $J$ is an interaction energy between nearest-neighbour spins $\langle i,j\rangle$,
and set to unity in our calculations. $s_g$ is the direction of ghost spin for the magnetization
of non-vanishing external field ($h>0$). For vanishing external field the model is
known to have a first-order PT. With increase of the external field, the first-order
PT line ends at a CP.

The order parameter of system is defined as,
\begin{flalign}\label{m}
m(T,L)=\frac{3\langle M\rangle}{2V}-\frac{1}{2}.
\end{flalign}
However, at CP, the original operators $E$ and $M$ lose their meaning as $T$-like and $H$-like, i.e., symmetry breaking couplings, as those in the Ising model. The order parameter and energy-like observable has to be redefined as the combination of the original $E$ and $M$, i.e.~\cite{potts-ising-b},
\begin{flalign}\label{M}
\tilde{M}=M+sE,\ {\rm and } \ \tilde{E}=E+rM.
\end{flalign}
The Hamiltonian in terms of $\tilde {M}$ and $\tilde {E}$ is,
\begin{flalign}\label{H}
H=\tau \tilde{E}-\zeta \tilde{M}.
\end{flalign}
Where new couplings are given by,
\begin{flalign}\label{eq-refer}
\zeta = \frac{1}{1-rs}(h-r \beta),\ {\rm and }\ \tau = \frac{1}{1-rs}(\beta-sh).
\end{flalign}
Here $r$ and $s$ are mixing parameters and have been determined in ref. ~\cite{potts-ising-b} by,
\begin{flalign}\label{r and s}
r^{-1}=(\frac{d \beta_{\rm C}(h)}{d h})_{h=h_{\rm C}},\ {\rm and} \ \langle\delta \tilde{M}\cdot \delta \tilde{E}\rangle =0,
\end{flalign}
with $\delta \tilde{X}=\tilde{X}-\langle\tilde{X}\rangle$ for $X=M$, or $E$.

The order parameter in terms of $\tau$ and $\zeta$ is,
\begin{flalign}\label{m and M}
m(\tau,\zeta)=\frac{1}{L^3}[\tilde M(\tau,\zeta)-\langle \tilde M(\tau_{\rm C},\zeta_{\rm C})\rangle].
\end{flalign}
It is the most sensitive observable to the PT. According to Eq.~(\ref{r and s}) and ~(\ref{m and M}), it can be written in terms of $T$ and $h$ as,
\begin{equation}
 m(T,h)= \frac{1}{L^3}[\tilde M(T,h)-\langle \tilde M(T_c,h_c)\rangle].
\end{equation}
Where $\tilde M(T,h)=M(T,h)+sE(T,h)$, and $M(T,h)$ is obtained by Eq.~(\ref{E and M}) from generated spins at lattice. The mixing parameter $s$ is estimated from the table 2 of ref.~\cite{potts-ising-b} and the Eq.~(\ref{r and s}).

In the following, we take the observable as the average absolute order parameter over all configurations, i.e.,
\begin{equation}
\hspace{-5.2mm}m(T,h)=
\langle|\frac{1}{L^3}[\tilde M(T,h)-\langle\tilde M(T_c,h_c)\rangle]|\rangle.
\end{equation}
\noindent For fixed external field $h$, it is the function of temperature and system size, i.e., $ m(T,L)$.

Three samples of order parameter are generated at three external fields, i.e., the first-order PT line ($h=0.0005$~\cite{Xue-JPG42}), the CP ($h=0.000775$~\cite{Kim-CEP,potts-ising-b}), and the crossover region ($h=0.002$~\cite{Xue-JPG42}), respectively. At each of external field, $18$ $T$-values are chosen starting from $T_0=1.8180$ with $\Delta T=0.0001$. For a pair of couplings $(\beta=1/T, h)$, 100,000 independent configurations are generated.

The simulation is performed by the Wolff cluster algorithm~\cite{wolff}, and the helical boundary conditions are used. Where a Ferrenberg-Swendsen reweighting analysis~\cite{multi-histogram} is used to calculate observables at intermediate parameter values. For each case, the simulation is carried out for four different lattice sizes $L=30,\ 40,\ 50,\ 60$.

The transition $T$ and $a$ for above three samples are listed in the bracket of Table I~\cite{Xue-JPG42, potts-ising-b}. They are obtained by conventional methods as mentioned in Introduction~\cite{Binder-a,Binder-b,sus-3}.

\section{Locating fixed points}

\begin{figure*}[t]
  \centering
  \noindent\makebox[\textwidth][c] {
    \includegraphics[width=0.85\paperwidth]{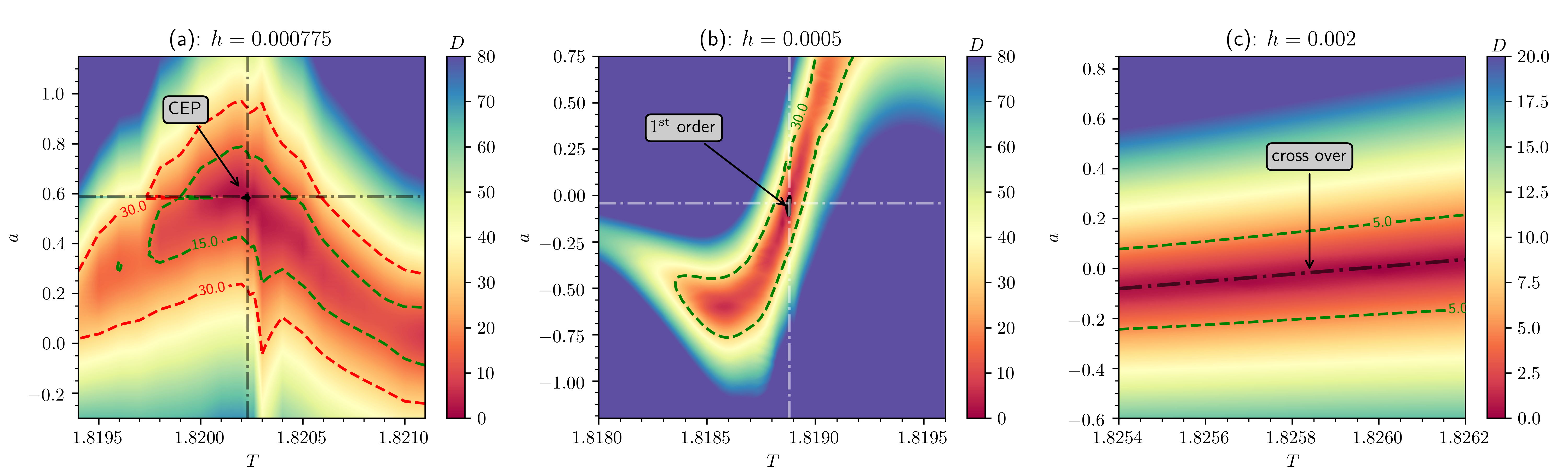} }
\caption{\label{Fig. 3} $D(T, a)$ for three samples with external fields $h=0.000775$ (a),  $0.0005$  (b),  and 0.002 (c). Dash-point lines label the coordinates of $D_{\rm min}(T, a)$ and dash lines are isolines.}
\end{figure*}

Now, we have three samples, where observables (mean of order parameters) for different $T$ and $L$ are presented. For each of samples, we calculate its corresponding width $D(T, a)$ and plot it in $T$ and $a$ plane.

\begin{table}
\centering
\setlength{\arrayrulewidth}{0.3pt}
\caption{Parameters of $T$ and $a$ at $D_{\rm min}(T, a)$ for three samples (Inside brackets are corresponding ones from conventional methods).}
\begin{tabular}{|c|c|c|c|}

\hline
\multirow{2}{*}{\diagbox{Sample}{Para.}}&  \multirow{2}{*}{$D_{min}(T,a)$}       &\multirow{2}{*}{$T$}      &\multirow{2}{*}{$a$} \\
                               &                                   &                                          &                               \\
 \hline
 \multirow{3}{*}{$2^{\rm nd}$ order PT}&\multirow{3}{*}{1.0291$\pm$0.2946} &\multirow{3}{*}{\minitab[c]{1.82023\\(1.82023372)}}&\multirow{3}{*}{\minitab[c]{0.583\\(0.564)}}\\
                               &                                                                 &                         &                      \\
                               &                                                                 &                         &\\

\hline
 \multirow{2}{*}{\minitab[c]{$1^{\rm st}$ order PT}}&\multirow{2}{*}{1.5287$\pm$0.5591}&\multirow{2}{*}{\minitab[c]{1.81887\\(1.8188763)}}&\multirow{2}{*}{\minitab[c]{-0.047\\(-0.0541)}}\\
                             &                                 &                             &       \\
\hline
 \multirow{3}{*}{\minitab[c]{crossover}}&\multirow{3}{*}{\minitab[c]{$\sim 1$ for all $T$}}&\multirow{3}{*}{1.8256-1.8261}&\multirow{3}{*}{-0.05$\sim$0.05}\\
                             &                                 &                             &       \\
                            &                                 &                             &       \\
\hline
\end{tabular}
\end{table}

According to the definition Eq.~(\ref{define D}), the width of order parameter is,
\begin{equation}
D(T,a)=\sqrt{\frac{\Delta S_{ m(T,L) L^{a}}}{N_L-1}}.
\end{equation}
Where,
\begin{equation}
   \begin{array}{cl}
 \displaystyle \hspace{-5.5mm}  \Delta S_{ m(T,L)L^{a}}=\sum_{i=1}^{N_L}\frac{1}{\omega^2_i}\times
       \big[m(T,L_i)L^{a}_i - \langle m(T,L)L^{a}\rangle\big]^2 .\\
    \end{array}
\end{equation}
$\omega_i$ is the error of  $m(T,L_i)L^{a}_i$. For a given lattice size $L$, $\omega_i$ mainly comes from $m(T,L_i)$, and is estimated by the square root of the variance of $m(T,L_i)$.
\begin{equation}
\langle m(T,L) L^{a}\rangle = \frac{\sum_{i=1}^{N_L} m(T,L_i)
L^{a}_i /\omega^2_i}{\sum_{i=1}^{N_L}1/\omega^2_i}
\end{equation}
is error weighted average. The summation number $N_L$ equals to 4 for four system sizes $L=30,\ 40,\ 50,\ 60$.

The contour plot of $D(T, a)$ in the plane of $T$ and $a$
for three samples are presented in Fig.~3(a), (b) and (c), respectively. Where the colour bars
on the right side of sub-figures indicate the values of $D(T, a)$. The range of $a$ is from $-1.2$ to $1.15$ with interval 0.05. The red and blue zones are minimum and maximum, respectively. The dash-point lines indicate the coordinates of $T$ and $a$ corresponding to the minimum of $D(T, a)$.

For the sample with external field $h=0.000775$, the contour lines in Fig. ~3(a) show that $D(T, a) $ gradually converges to a minimum, the red area. This means that the width converges to a minimum at a specified $T$ and $a$. It is typical features of FP.

\begin{figure}[hbt]
\centering
\includegraphics[width=3.5in]{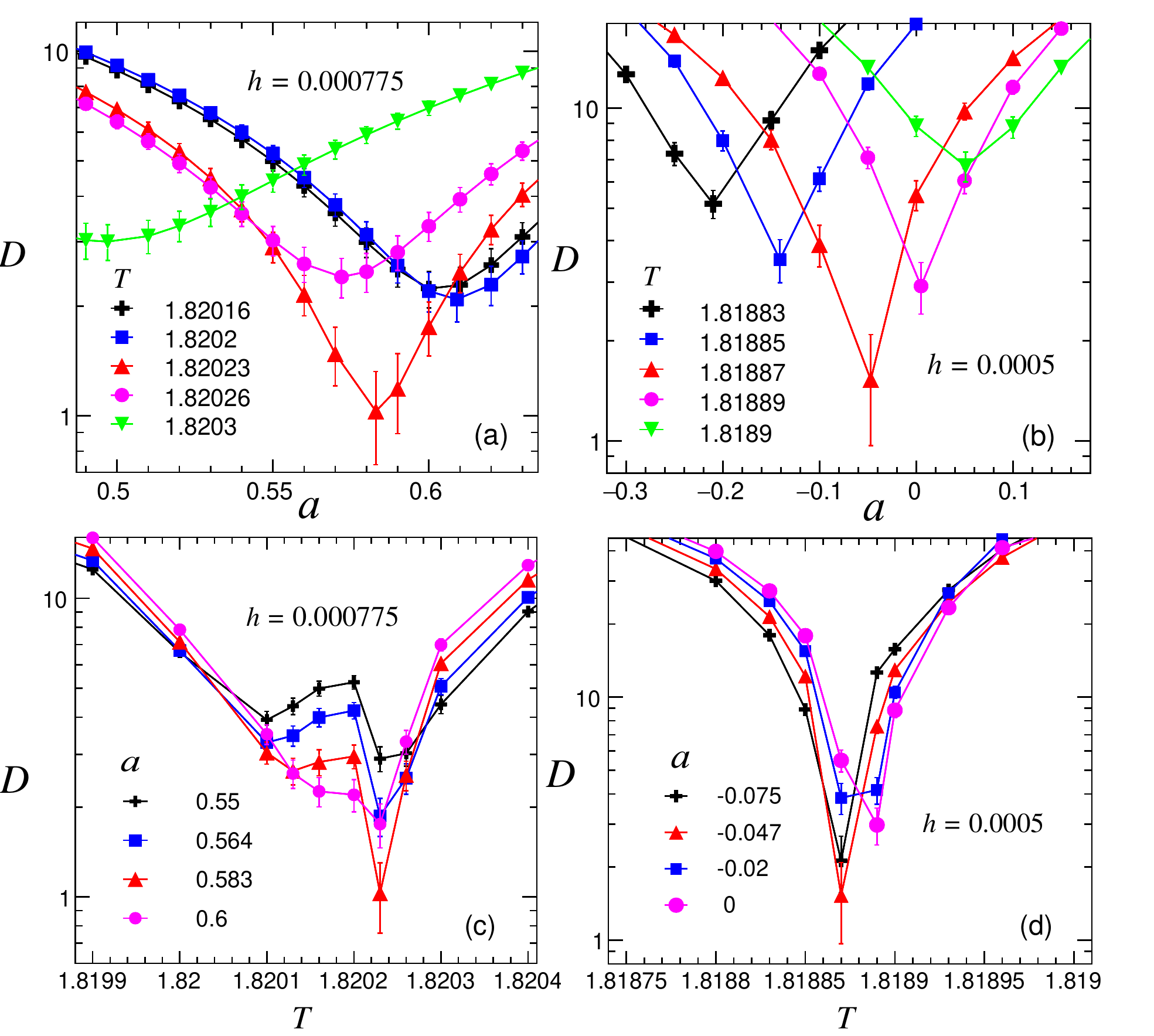}
\caption{\label{Fig. 4} Projections of $D(T, a)$ on exponent ratio (upper row) and temperature (lower row) axes nearby the critical point (a) and (c), and the point of the first order phase
transition line (b) and (d). }
\end{figure}

To amplify the fine structure and minimum nearby the red area, we project
$D(T, a)$ to $a$ and $T$ axes, respectively, as showed in Fig.~4(a) and 4(c). In Fig.~4(a),
for a given $T$, there is an $a$ which makes $D(T, a)$ minimum. Among these lines,
the minimum of the red line is the smallest one, where $D_{\rm min}=1.0291\pm  0.2946$, and corresponding $T=1.82023$ and $a=0.583$. They are very close to $1.82023372$ and
$0.564$, as listed in the blankets of the second row of Table I.

From the projection along the direction of temperature as showed in Fig.~4(c), for a given
ratio $a$, there is also a minimum $D(T, a)$. The minimum of the red line is the smallest one among them. Its corresponding $T$ and $a$ are the same as those from Fig.~4(a).

These features of $D(T, a)$ and value of $a$ are consistent with the FP of the second-order PT, i.e., CP, as showed in the case of II in Fig.~2.

Turn to the sample with external field $h=0.0005$, the contour lines in
Fig.~3(b) also shows that $D(T, a)$ gradually converges to a minimum, the red area. This means that all curves of scaled observable for different sizes are more and more close to each other with the change of $T$ and $a$, as expected for FP.

The projection plots of $D(T, a)$ versus $a$ for different $T$ are showed in
Fig.~4(b). The curve which has the smallest minimum is the red one with
$ D_{\rm min}(T, a)=1.5287 \pm 0.5591 $, $T=1.81887$, and $a=-0.047$.
They are very close to the $T=1.8188763$, and $a=-0.0541$, as listed in the blankets of the third row of Table I. Here the order parameter can be considered as the first-order of susceptibility, i.e., $n=1$. So $a=(n-1)d=0$.

The projection along the direction of $T$ is showed in Fig.~4(d), for a given
ratio $a$, there is also a minimum $D(T, a)$. The smallest minimum gives the same $T$ and $a$ as those from Fig.~4(b). Those are the features of the FP of the first-order PT line, as showed in the case of I in Fig.~2.

For the sample with external field $h=0.002$, the contour lines of $D(T, a)$ in Fig.~3(c) are bands  parallel to the $T$-axis, in contrast to Fig.~3(a) and 3(b).
This implies that $D(T, a)$ is independent of $T$. Its value is determined by $a$ only.
The minimum, the red band, is close to zero. These are the features of the FP of the crossover region,
as showed in the case of III in Fig.~2.

These three examples show that the defined width is very sensitive to $T$ and $a$. The minimum of $D(T, a)$ precisely determines the $T$ and $a$ of FP. Therefore, as long as FP exists in the $T$ and $a$ plane, the minimum of defined width can clearly locate it.

The contour area of defined width $D(T,a)$ as showed in Fig.~3 in fact indicates critical or transitional ranges of temperature $T$ and critical exponent ratio $a$, where all curves for different sizes start to converge toward each other. Out of the area, $D(T,a)$ is big and uniformly blue, i.e., all curves are apart from each other. This contour area for the second order PT as showed in Fig.~3(a) is obviously larger than that of the first order PT as showed in Fig.~3(b), where the temperature range is much narrow. Therefore, the PT of the first order is very sensitive to the change of $T$.

In critical or transitional range, for a given temperature $T$ ($a$), there is a $a$ ($T$) made the $D(T,a)$ minimum, as showed in Fig.~4. This minimum implies that all curves for different sizes are closest to each other at given $T$ ($a$). Usually this minimum is larger than unity, i.e., all curves are still not overlapped within error. Only for the critical or transitional $T$ ($a$), the minimum of $D(T,a)$ is the smallest and around unity, i.e., all curves converge to a FP within errors. So defined width well quantifies the behaviour of FP, the ranges of critical or transitional $T$ and $a$, and the deviation of $T$ and $a$ from the ranges.

On the other hand, if some system sizes are too small to the FSS, their corresponding curves in the plot like
Fig.~1(b) or (d) will not converge exactly to the FP within error. In the case, the smallest minimum of $D(T,a)$ will be larger than unity. Therefore, $D(T,a)$ is around unity if and only if various of system sizes are all properly large and all parameters in the FSS Eq.~(\ref{scaling function 1}) are critical or transitional values.

Certainly, a real sample may not be so well presented as above three samples. It depends on the observable, the covered area of phase plane, and experimental settings.

Firstly, as we know, some observable, such as energy density and specific heat may not exactly follow the FSS in the vicinity of $T_{\rm C}$~\cite{Book-phase}. For this kind of observable, additional terms of scaling violation are not negligible~\cite{Engels:2011km, Bloete}. Scaling function usually varies with the system size and temperature. There is no FP~\cite{Engels:2011km, Bloete}.

Secondly, the suggested contour plot depends on its covered area of phase plane. If it is far away from the phase boundary, the observable is independent either of temperature or system size~\cite{Xue-JPG42}, and the plot is constantly big blue, as showing in Fig.~3. If it approaches to the phase boundary, or the transition temperature, the plot may show some contour areas which are similar to a part of fig.~3(a), or (b), or (c). Therefore, the contour plot of defined width is helpful in exploring FPs of the phase boundary.

It should also be noticed that due to the error of observable and uncertainties of related parameters in real experimental settings, the minimum of $D(T, a)$ may be larger than the unity and vary with experiments, what's more important for the formation of a FP is the trend of $D(T, a)$ converging to a minimum.

\section{A possible application}

One of the main goals of contemporary beam energy scan (BES) at relativistic heavy-ion collisions (RHIC) and future CBM/FAIR and NICA experiments is to study the critical end point (CEP) in the QCD phase diagram in terms of existence and location~\cite{scan-c,STAR-BES, STAR-c, Fisch-QCD, Luo-Xu-Nature,NICA,FAIR}. As incident energy of the collision ($\sqrt{s_{NN}}$) changes, temperature $T$ and baryon chemical potential $\mu_B$ of formed system in the phase plane change accordingly. How they change is directed by the simplification of thermal model descriptions of the freeze-out region~\cite{obversables-scaling-b,center-energy,S-T-u}. Therefore, the purpose of the BES program is to scan the observable in the $T$ and $\mu_B$ plane.

Due to the small volume and the short lifetime of the quark-gluon plasma (QGP) formed in heavy ion collisions, the CEP, if appears, will be blurred. Critical fluctuations are severely influenced by the finite volume, as well as finite-time. Due to finite evolution time and critical slowing down, the system may not reach thermal equilibrium, or pass the CEP~\cite{correlation,slowing-down}. The non-equilibrium evolution depends on the given dynamics which are currently unclear, and should be examined with caution.

For a restricted volume which is not very small, the singularity of generalized susceptibility $\chi_i$  is limited into a finite peak with modified position and width~\cite{finite-size-a,finite-size-b}. With the decrease of volume size, the peak position shifts towards lower temperature and larger chemical potential~\cite{brazil, PRD90-2014, PLB713-2012}. It is the so called pseudo-critical point. The precise position of CEP has to be determined by the FSS of observable~\cite{Wu-Lizhu,Lizhu-wu,obversables-scaling-b}.

The FSS of several observables in relativistic heavy-ion collisions have been studied~\cite{obversables-scaling-b, RoyPRL, Roy1, Roy2, PRD97-NA49}. In this paper, we suggest locating FP by the newly defined width of all points of scaled observables for different sizes at a given $T$ and $a$. The minimum of the width in $T$ and $a$ plane is the place of FP. Graphing with the width allows us to easily locate the FP (should it exist in the phase plane) without selecting the best scaling plot of observable, thus eliminating inaccuracies inflicted by human observation. The value of $a_{\rm FP}$ tells either the FP is a CP, a point of the first-order PT line, or a point in the crossover region. So classifying the observed FP becomes possible.

As we know, there are still numbers of uncertainties in applying the FSS in relativistic heavy-ion collisions~\cite{obversables-scaling-b, RoyPRL}: whether the observable is properly chosen, whether the phase boundary is covered by the BES, and whether the system size can be correctly estimated.

The system size is related to the given centrality of heavy-ion collisions. It may be influenced by the incident energy as well. The relations between them have not been set up quantitatively, and should be studied carefully in the future. Currently, the radii of Hanbury Brown Twiss (HBT) interferometry only provides a rough estimation~\cite{HBT-1, HBT-2, HBT-3, RoyPRL}.

In particular, it is difficult to define an appropriate observable in heavy-ion collisions. Although the order parameter of chiral and deconfining PT is well defined from the theoretical side~\cite{Fisch-QCD}, its corresponding observable is still not known. Certainly, it is better to measure observables, like the order parameter in the Potts model, or the susceptibility, as discussed at the end of the above section. However, it is impossible to know in prior if the observable is analogous with either the order parameter, the susceptibility, or the specific heat.

Nevertheless, it should be helpful and worthwhile to try the method by suggested observables, such as event-by-event fluctuations of the net baryon number, the electric charge, or the strangeness of the heavy ion system~\cite{cumulant-lattice-b,cumulant-lattice-a,cumulant-b}, and explore the possible FP at the RHIC BES as was delineated in refs~\cite{obversables-scaling-b, RoyPRL}.

\section{Summary and conclusions}

In the finite size scaling, the critical point is a fixed point, where all scaled observables of different system sizes intersect. Fixed point also exists on the first-order PT line and can be generalized to the crossover region. Their corresponding scaling exponent ratios are fraction, integer and zero, respectively.

To quantify the feature of FP, we define, at a given $T$ and $a$, the width of scaled observables for different sizes. When $T=T_{\rm FP}$ and $a=a_{\rm FP}$, the defined width reaches its minimum.

Then, using the 3D three-state Potts model, we produce three samples at three external fields. These three samples contain, respectively, the CP, the point of the first-order PT line, and the point of the crossover region. The temperature covers the whole possible range. We calculate the width of order parameters in all three samples, and plot them in the $T$ and $a$ plane.

The contour plot of defined width clearly shows the critical or transitional ranges temperature and critical exponent ratio. The minimum of defined width locates precisely the position of FP in three samples, respectively. These demonstrate the method is effective and precise in locating all three categories of FP.
The defined width well quantifies the behaviour of FP.

Therefore, from simply scanning the observable in the phase plane, FP can be well located by defined width. Finally, a possible application at the RHIC BES is discussed. It should be helpful in locating the FP of QCD PT from experimental side.

\section{Acknowledgement}

We are grateful to Dr. Xiaosong Chen for drawing our attention to the field.

This work is supported in part by the Major State Basic Research Development Program of China under Grant No. 2014CB845402, the Ministry of Science and Technology (MoST) under grant No. 2016YFE0104800.

%

\ed